# Approximating Minimum-Cost $k$-Node Connected Subgraphs via Independence-Free Graphs

Joseph Cheriyan[*]    László A. Végh[†]

November 10, 2018

**Abstract**

We present a 6-approximation algorithm for the minimum-cost $k$-node connected spanning subgraph problem, assuming that the number of nodes is at least $k^3(k-1)+k$. We apply a combinatorial preprocessing, based on the Frank-Tardos algorithm for $k$-outconnectivity, to transform any input into an instance such that the iterative rounding method gives a 2-approximation guarantee. This is the first constant-factor approximation algorithm even in the asymptotic setting of the problem, that is, the restriction to instances where the number of nodes is lower bounded by a function of $k$.

## 1 Introduction

A basic problem in network design is to find a minimum-cost sub-network $H$ of a given network $G$ such that $H$ satisfies some prespecified connectivity requirements. Most of these problems are NP-hard, hence, research has focused on the design and analysis of approximation algorithms. The area flourished in the 1990s, and there were a number of landmark results pertaining to problems with edge-connectivity requirements. This line of research culminated in a result of Jain that gives a 2-approximation algorithm for a general problem called the *survivable network design problem*, abbreviated as SNDP.[1] Progress has been much slower on similar problems with node-connectivity requirements, despite more than a decade of active research.

Our focus is on undirected graphs throughout. For a positive integer $k$, a graph is called *k-node-connected* (abbreviated *k-connected*) if it has at least $k+1$ nodes, and the deletion of any set of $k-1$ nodes leaves a connected graph. In the *minimum-cost k-connected spanning subgraph* problem, we are given a graph with nonnegative costs on the edges; the goal is to find a $k$-connected spanning subgraph of minimum cost. Let $opt(G)$ denote the cost of an optimal subgraph. Throughout, we use $k$ to denote the connectivity parameter, and $n = |V|$ to denote the number of nodes; both are integers with $1 \leq k < n$.

---

[*]Dept. of Combinatorics & Optimization, University of Waterloo, Waterloo, Ontario N2L3G1, Canada. (jcheriyan@uwaterloo.ca) Supported by NSERC grant No. OGP0138432.

[†]Dept. of Management, London School of Economics & Political Science, Houghton Street, London WC2A 2AE, UK. (l.vegh@lse.ac.uk)

[1] In the SNDP, we are given an undirected graph with non-negative costs on the edges, and for every unordered pair of nodes $i, j$, we are given a number $\rho_{i,j}$; the goal is to find a subgraph of minimum cost that has at least $\rho_{i,j}$ edge-disjoint paths between $i$ and $j$ for every pair of nodes $i, j$.

## 1.1 Previous results

The problem is NP-hard for $k \geq 2$. It is easy to obtain an approximation guarantee of $2k$, by applying the result of Frank and Tardos [9] on $k$-outconnectivity; this is discussed in [17]. This approximation guarantee was improved to $k$ by Kortsarz and Nutov [18].

In the *asymptotic setting* of the problem, we restrict ourselves to instances where the number of nodes is lower bounded by a function of $k$. Results in the asymptotic setting address the issue of approximability as a function of the single parameter $k$ (for all sufficiently large $n$). In [3], an $O(\log k)$ approximation guarantee was given for the asymptotic setting, assuming that $n \geq 6k^2$.

Most research efforts subsequent to [3] focused on finding near-logarithmic approximation guarantees for all possible ranges of $n$ and $k$, and on extending the results to the more general setting of directed graphs. Kortsarz and Nutov [19] presented an algorithm with an approximation guarantee of $O(\log k \cdot \min\{\sqrt{k}, \frac{n}{n-k} \log k\})$. The paper by Fakcharoenphol and Laekhanukit [4] gave an $O(\log^2 k)$-approximation algorithm. The approximation guarantee was further improved by Nutov [25] to $O(\log k \log \frac{n}{n-k})$. The results of [19, 4, 25] apply to both undirected graphs and directed graphs. The approximability for $k = n - o(n)$ seems to raise combinatorial difficulties such that even a decade after the $O(\log k)$ approximation guarantee was proved in the asymptotic setting, it is still not clear whether the same guarantee holds for all $k$ and $n$.

Even the following fundamental question has been open: Does there exist an $o(\log k)$ approximation algorithm for the problem on undirected graphs in the asymptotic setting, or is it possible to prove a hardness-of-approximation threshold strictly larger than $\Omega(1)$?

## 1.2 Our result and the main ideas

We resolve the above question by proving the following result.

**Theorem 1.1.** *Let $G = (V, E)$ be an undirected, $k$-connected graph with at least $k^3(k-1) + k$ nodes. There is a polynomial-time algorithm that finds a $k$-connected spanning subgraph of cost $\leq 6\mathrm{opt}(G)$.*

In what follows, we describe the main ideas of our result. Whereas no constant factor approximation was given previously for this problem, such results were already known for similar problems with edge-connectivity requirements. Jain [15] introduced the iterative rounding method (see Figure 1), and used it to design a 2-approximation algorithm for the SNDP. Jain's pivotal result asserts that every basic feasible solution to the standard linear programming (LP) relaxation has at least one edge of value at least $\frac{1}{2}$. A 2-approximation is obtained by iteratively adding such an edge to the graph and solving the LP relaxation again.

As tempting as it might be to apply iterative rounding for SNDP with node-connectivity requirements, unfortunately the standard LP relaxation for this problem might have basic feasible solutions with small fractional values on every edge. Such examples were presented in [2, 5, 6]. Recently, [1] improved on these previous constructions[2] by exhibiting an example of the min-cost $k$-connected spanning subgraph problem with a basic feasible solution that has value $O(1/\sqrt{k})$ on every edge. Still, iterative rounding has been applied to problems with node-connectivity requirements: Fleischer, Jain and Williamson [6] gave a 2-approximation for node-connectivity SNDP with maximum requirement 2. Recently, Nutov [26], and Fukunaga and Ravi [11] studied degree-bounded variants of node-connectivity SNDP and gave bicriteria approximations based on iterative rounding.

---
[2] The construction in [1] applies to our problem, whereas the negative implications of the constructions predating [1] apply to more general problems (e.g., node-connectivity SNDP) but not to our setting.

Our new insight is that whereas iterative rounding fails to give $O(1)$-approximations for arbitrary instances, we can isolate a class of graphs where it does give $O(1)$-approximations; and moreover, we are able to transform an arbitrary input instance to a new instance from this class.

## 1.3 Independence-free graphs

There is an equivalent formulation of our problem that we prefer to use within this paper: For a set $V$, let $\binom{V}{2}$ denote the edge set of the complete graph on the node set $V$. In the *minimum-cost $k$-connectivity augmentation problem*, we are given a graph $G = (V, E)$ and nonnegative edge costs $c : \binom{V}{2} \to \mathbb{R}_+$, and the task is to find a minimum cost set $F \subseteq \binom{V}{2}$ of edges such that $G + F$ is $k$-connected.[3] Let $\mathrm{opt}(G)$ denote the cost of an optimal augmenting edge set. Our reason for switching problems is the formal convenience of the connectivity augmentation framework for the presentation of iterative rounding as the second part of our algorithm; the standard analysis of iterative rounding is "memoryless" in that the analysis holds regardless of the "starting graph", whereas our analysis of iterative rounding exploits properties of this graph.

We show that the failure of the iterative rounding method for node-connectivity requirements can be attributed to a specific structure, that we now informally describe; Section 2 contains the definitions and details. Frank and Jordán [8] introduced the framework of set-pairs for node-connectivity problems; the LP relaxation is also based on this notion. By a *set-pair*, we mean a pair of nonempty disjoint sets of nodes, not connected by any edge of the graph; the two sets are called *pieces*. If the union of the two pieces has size $> n - k$, then the set-pair is called *deficient*, since it corresponds to the two sides of a node cut of size $< k$. Clearly, a $k$-connected graph must not contain any deficient set-pairs. A new edge has to cover every deficient set-pair, that is, an edge whose endpoints lie in the two different pieces. Two set-pairs are called *dependent*, if they can be simultaneously covered by an edge (of the complete graph), otherwise, the two set-pairs are called *independent*. It can be seen that the two set-pairs are independent if and only if one of them has a piece disjoint from both pieces of the other set-pair; see Figure 2(a) for an illustration of such a configuration. A graph is called *independence-free* if any two deficient set-pairs are dependent. We observed that bad examples for iterative rounding (such as the one in [1]) always contain independent deficient set-pairs. We show that this is the only possible obstruction: in independence-free graphs, the analog of Jain's theorem holds, that is, every basic feasible solution to the LP relaxation has an edge with value at least $\frac{1}{2}$; see Theorem 2.1 in Section 2.

Theorem 2.1 can be essentially derived from a general result by Fleischer et al. [6, Theorem 3.5], asserting that iterative rounding gives a 2-approximation for covering weakly two-supermodular functions. This is an extension of Jain's notion of weakly supermodular (requirement) functions to the framework of set-pairs; Fleischer et al. showed that Jain's token arguments can be extended to derive the result. We include a direct, simpler proof of Theorem 2.1, using the fractional token argument of Nagarajan et al. [24].

---

[3] Let us quickly verify the equivalence of the two problems. Given an instance $(V, \hat{E})$, $\hat{c} : \hat{E} \to \mathbb{R}_+$ of the subgraph problem, we can reduce it to the augmentation problem with $G = (V, \emptyset)$, $c_e = \hat{c}_e$ if $e \in \hat{E}$ and $c_e = \infty$ if $e \in \binom{V}{2} - \hat{E}$. In the other direction, given an instance $G = (V, E)$, $c : \binom{V}{2} \to \mathbb{R}_+$ of the augmentation problem, we can reduce it to the subgraph problem on the complete graph, with $\hat{c}_e = c_e$ if $e \in \binom{V}{2} - E$ and $\hat{c}_e = 0$ if $e \in E$. Note that parallel edges are not relevant in both problems, that is, any solution subgraph can be assumed to be a simple graph.

The notion of independence-free graphs was introduced by Jackson and Jordán [14] in the context of minimum cardinality $k$-connectivity augmentation (the special case of our problem where each edge in $\binom{V}{2} - E$ has cost 1). They gave a polynomial-time algorithm for this problem for fixed $k$. They first solve the problem for independence-free graphs and then show how the general case can be reduced to such instances. At a high level, we follow a similar approach, but there is very little in common between the details of their algorithm and ours; they have to use an elaborate analysis to get an optimal solution to an unweighted problem, whereas we use simple methods (based on powerful algorithmic tools) to approximately solve the weighted problem. The first phase of our algorithm uses "combinatorial methods" to add a set of edges of cost $\leq 4\text{opt}(G)$ to obtain an independence-free graph. The second phase of our algorithm then applies iterative rounding to add a set of edges of cost $\leq 2\text{opt}(G)$ to obtain an augmented graph that is $k$-connected.

## 1.4 Overview of the first phase

In the first phase, we shall guarantee a property stronger than independence-freeness. For this purpose, let us consider deficient sets instead of deficient set-pairs. A set of nodes $U$ is called deficient, if it has less than $k$ neighbours, and moreover, the union of $U$ and its neighbour-set is a proper subset of $V$ (in other words, the neighbours of $U$ form a node cut of size $< k$). There is a one-to-one correspondence between deficient sets and pieces of deficient set-pairs. By a *rogue set* we mean a deficient set $U$ with $|U| < k$. We call a graph *rogue-free* if it does not contain any rogue-sets; or equivalently, if every deficient set is of size at least $k$. It is easy to see that a rogue-free graph must also be independence-free.

Next, we give an algorithmic overview of the first phase by showing that an arbitrary graph $G$ with at least $k^3(k-1) + k$ nodes can be made rogue-free by two applications of the Frank-Tardos algorithm [9] for $k$-outconnectivity. (Section 3.1 discusses this algorithm in sufficient detail; it is a standard tool in the area, and has been used in [17, 3, 19, 4, 25], etc.) First, we pick a set $R_0$ of $k$ arbitrary nodes of $G$ and connect them (temporarily) to a new root node $r$. Then we apply the Frank-Tardos algorithm with root $r$; after recording the output, we remove $r$ and its incident edges. The algorithm outputs a set of edges $F_0$ of cost $\leq 2\text{opt}(G)$ such that in the augmented graph $G' = G + F_0$, every surviving deficient set contains some node of $R_0$. Theorem 2.4 below asserts that the union of all rogue sets of $G'$ has size $\leq k^3(k-1)$. In Section 5, assuming that $n \geq k^3(k-1) + k$, we describe a polynomial-time algorithm for finding (a superset of) the union of rogue sets. Hence, we can choose a second set of nodes $R_1$ of size $k$, disjoint from all rogue sets, and apply the Frank-Tardos algorithm again to find a set of edges $F_1$ of cost $\leq 2\text{opt}(G)$ such that in the augmented graph $G'' = G' + F_1 = G + F_0 + F_1$, every surviving deficient set contains some node of $R_1$. The key point is that the graph $G''$ resulting from the second application has no rogue sets (any rogue set of $G''$ must be a rogue set of $G' = G'' - F_1$, and moreover, it must contain a node of $R_1$, but we chose $R_1$ to be disjoint from all rogue sets of $G'$). Thus, we make the graph independence-free by adding a set of edges of total cost $\leq 4\text{opt}(G)$.

We restate our main result in the setting of the min-cost $k$-connectivity augmentation problem.

**Theorem 1.2.** *Let $G = (V, E)$ be an undirected graph with at least $k^3(k-1) + k$ nodes. There is a polynomial-time algorithm that finds an edge set $F \subseteq \binom{V}{2}$ such that $G + F$ is $k$-connected and*

$c(F) \leq 6\mathrm{opt}(G)$.

The rest of the paper is organized as follows. Section 2 precisely defines the notion of set-pairs, the LP relaxation, independence-free and rogue-free graphs, and formulates the two main theorems of the two parts of the proof. Section 3 bounds the size of the union of the rogue sets after the first application of the Frank-Tardos algorithm. Section 4 analyses the iterative rounding method on independence-free graphs. The arguments of these sections do not rely on each other. Section 5 is devoted to algorithmic aspects. Whereas we show in Section 3 that the union of rogue-sets has size $\leq k^3(k-1)$, we are not able to find this union efficiently; here we show how to circumvent this problem. Finally, Section 6 discusses some related problems and open questions.

## 2 Set-pairs, LP relaxation, and independence

For a set $U \subseteq V$, we use $\Gamma(U)$ to denote the set of neighbours of $U$, namely, $\{w \in V - U \mid \exists uw \in E, u \in U\}$, and we use $\gamma(U)$ to denote $|\Gamma(U)|$. Let $U^* = V - (U \cup \Gamma(U))$. By a *deficient set* $U$ we mean a set of nodes $U$ such that $\gamma(U) < k$ and $U$ and $U^*$ are both nonempty. Clearly, a graph is $k$-connected if and only if there are no deficient sets in it.

A more abstract yet more convenient characterization of $k$-connectivity can be given in terms of set-pairs. Note that set-pairs are usually defined in a directed sense, see [8, 3]. Since our focus is on undirected graphs, our set-pairs are defined as unordered pairs.

For two disjoint nonempty sets of nodes $U_0$ and $U_1$, the unordered pair $\mathbb{U} = (U_0, U_1)$ is called a *set-pair* if there is no edge with one end in $U_0$ and the other end in $U_1$. $U_0$ and $U_1$ are called the *pieces* of $\mathbb{U}$. We use $\Gamma(\mathbb{U}) = \Gamma(U_0, U_1)$ to denote $V - (U_0 \cup U_1)$. Let us define the deficiency function

$$p(\mathbb{U}) = p(U_0, U_1) = \max\{0, k - |\Gamma(U)|\} = \\ \max\{0, k - |V - (U_0 \cup U_1)|\}. \qquad (1)$$

The set-pair is called *deficient* if $p(\mathbb{U}) > 0$. It is easy to see that a graph is $k$-connected if and only if there are no deficient set-pairs, that is, $p \equiv 0$. Furthermore, if $U$ is a deficient set, then $(U, U^*)$ is a deficient set-pair with $\Gamma(U) = \Gamma(U, U^*)$ and $p(U, U^*) = k - \gamma(U) > 0$. Conversely, if $(U_0, U_1)$ is a deficient set-pair, then both $U_0$ and $U_1$ are deficient sets with $U_0 \subseteq U_1^*$ and $U_1 \subseteq U_0^*$.

We say that an edge $e = uv \in \binom{V}{2}$ *covers* the set-pair $\mathbb{U} = (U_0, U_1)$, if one of its endpoints lies in $U_0$ and the other one lies in $U_1$. For an edge set $F \subseteq \binom{V}{2}$, let $d_F(\mathbb{U}) = d_F(U_0, U_1)$ denote the number of edges in $F$ covering $\mathbb{U}$. Clearly, the following statement holds: $G + F$ is $k$-connected if and only if $d_F(\mathbb{U}) \geq p(\mathbb{U})$ for every set-pair $\mathbb{U}$.

Let $\mathcal{S}$ denote the family of all set-pairs in $G$, and for a set-pair $\mathbb{U}$, let $\delta(\mathbb{U}) \subseteq \binom{V}{2}$ denote the set of edges covering $\mathbb{U}$. For a vector $x : E \to \mathbb{R}$ and a set-pair $\mathbb{U}$, let $x(\delta(\mathbb{U})) = \sum_{e \in \delta(\mathbb{U})} x_e$. The following is a well-known LP relaxation of the minimum cost $k$-connectivity augmentation problem.

$$\begin{aligned} \text{minimize} \quad & \sum_{e \in E} c_e x_e & & \text{(LP-VC)} \\ \text{subject to} \quad & x(\delta(\mathbb{U})) \geq p(\mathbb{U}), \quad \forall \mathbb{U} \in \mathcal{S} \\ & x_e \geq 0, \quad \forall e \in \binom{V}{2} \end{aligned}$$

Requiring integrality of the variables $x_e$ we get the integer programming formulation of the problem. Notice that an optimal integral solution contains neither any edge of the original graph $G$ nor any parallel edges.

We say that two set-pairs $\mathbb{U} = (U_0, U_1)$ and $\mathbb{W} = (W_0, W_1)$ are *independent* if there is no edge in $\binom{V}{2}$ covering both of them. It is easy to verify that $\mathbb{U}$ and $\mathbb{W}$ are independent if and only if either $\mathbb{U}$ has a piece disjoint from both pieces of $\mathbb{W}$, or $\mathbb{W}$ has a piece disjoint from both pieces of $\mathbb{U}$.

The graph $G = (V, E)$ is called *independence-free* if it does not have two set-pairs that are deficient and independent; in other words, for every two deficient set-pairs $\mathbb{U} = (U_0, U_1)$ and $\mathbb{W} = (W_0, W_1)$, there exists $i \in \{0, 1\}$ such that $U_0$ intersects $W_i$ and $U_1$ intersects $W_{1-i}$.

We shall prove the following result in Section 4. As mentioned in the Introduction, it can also be derived via a slight modification of the proof of Theorem 3.5 in Fleischer et al. [6].[4]

**Theorem 2.1.** *Let $G = (V, E)$ be an independence-free graph and let $k$ be a positive integer. Then every basic feasible solution $x$ to (LP-VC) has an edge $e$ with $x_e \geq 1/2$.*

Iterative rounding was introduced by Jain [15] for survivable network design; we refer the reader to the recent book [21] on this method. It can be naturally adapted to our problem of min-cost $k$-connectivity augmentation, as outlined in Figure 1. The next corollary follows directly from Theorem 2.1, using the standard argument from [15]; observe that adding new edges to an independence-free graph preserves this property.

**Corollary 2.2.** *The iterative rounding algorithm in Figure 1 returns an edge set of cost $\leq 2\mathrm{opt}(G)$.*

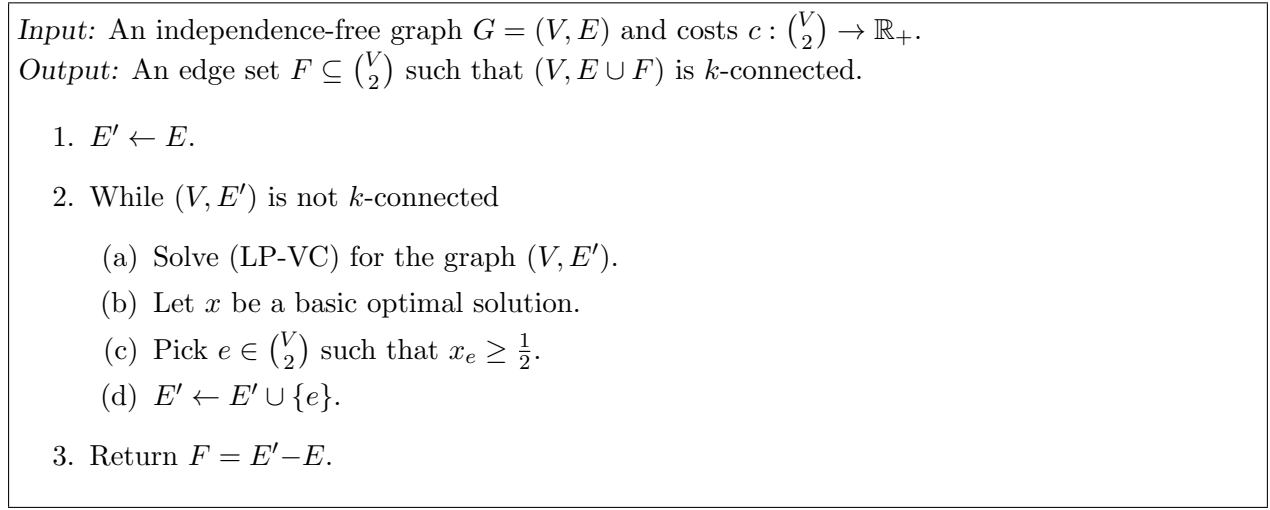

Figure 1: Iterative rounding algorithm

We call a deficient set $U$ with $|U| < k$ a *rogue set*. A graph is called *rogue-free* if there are no rogue sets in it, that is, every deficient set is of cardinality $\geq k$. Whenever we have two set-pairs $(U_0, U_1)$ and $(W_0, W_1)$ that are independent, then at least one of the four pieces, $U_0, U_1, W_0, W_1$ must be a rogue set. We state this for later use.

---
[4]In fact, the connectivity requirement function of independence-free graphs does not satisfy the weakly two-supermodular property of [6], but the property holds if we only require it for two set-pairs both with positive deficiency. It can be seen that the proof in [6] can be extended assuming this weaker property.

**Fact 2.3.** *If a graph has two set-pairs that are independent, then it has a rogue set. Equivalently, if a graph is rogue-free, then it is independence-free.*

Our main structural result on rogue sets follows. This result is the key to our first algorithmic goal, namely, given the input graph $G = (V, E)$, find an edge set $F_0$ such that $G + F_0$ is independence-free and $c(F_0) \leq 4\text{opt}$.

**Theorem 2.4.** *Assume that there exists a set $R \subseteq V$ such that every rogue set has a nonempty intersection with $R$. Then the union of all rogue sets has size $\leq |R|k^2(k-1)$.*

## 3 Making a graph rogue-free

In this section, we first describe our main algorithmic tool, the Frank-Tardos algorithm, and its use in the first phase of our algorithm. Section 3.2 is devoted to the proof of Theorem 2.4.

### 3.1 The Frank-Tardos algorithm for $k$-outconnectivity

Let $D = (V, E)$ be a directed graph, let $r$ be a node of $D$, and let $k$ be a positive integer; $D$ is called *$k$-outconnected* from $r$ (or, $k$-outconnected with root $r$) if it has $k$ internally disjoint dipaths from $r$ to $v$, for each node $v \in V - \{r\}$. This notion applies to undirected graphs too, provided we require internally disjoint paths rather than internally disjoint dipaths. Frank and Tardos [9] gave a polynomial-time algorithm for finding an optimal solution to the following problem: Given a directed graph $D$ with costs on the edges, a root node $r$, and a positive integer $k$, find a min-cost subgraph of $D$ that is $k$-outconnected from $r$. (See also Frank [7].)

We also refer to the analogous problem of *undirected $k$-outconnectivity augmentation*: Given a graph $G = (V, E)$, $r \in V$, a positive integer $k$, and non-negative costs on $\binom{V}{2}$, find a min-cost set of edges $F$ such that $G + F$ is $k$-outconnected from $r$.

In contrast to its directed counterpart, this problem is NP-hard for $k \geq 2$.[5] Fortunately, the result of Frank and Tardos immediately yields a 2-approximation algorithm for undirected graphs: we obtain a directed graph by replacing each edge $e$ of the graph by two oppositely oriented edges with the same cost as $e$, and then we apply the algorithm of [9] to this directed graph.

We shall apply this algorithm in the following special way. In the graph $G = (V, E)$, pick a set of nodes $R \subseteq V$, with $|R| = k$. By a *terminal* we mean a node of $R$. We (temporarily) add a new node $\hat{r}$ to the graph, and we define costs for the $n$ edges $\hat{r}v, v \in V$, by fixing $c_{\hat{r}v} = 0$ if $v \in R$, and $c_{\hat{r}v} = \infty$ if $v \in V - R$. Let $\hat{E}$ denote the set of edges between $\hat{r}$ and the terminals in $R$, and let $\hat{G} = (V + \hat{r}, E + \hat{E})$ denote the resulting graph. Note that if $G + F$ is $k$-connected, then $\hat{G} + F$ is $k$-outconnected with root $\hat{r}$. Consequently, the optimal cost of $k$-outconnectivity augmentation is $\leq \text{opt}(G)$. We apply the Frank-Tardos algorithm to find a set of edges $F'$ of cost $\leq 2\text{opt}(G)$ such that $\hat{G} + F'$ is $k$-outconnected from $\hat{r}$; note that the cost of each edge of $F'$ incident to $\hat{r}$ is zero. Finally, we remove the root $\hat{r}$ and all edges incident to it. We refer to this procedure as *$R$-outconnectivity augmentation*, and we denote it as subroutine $\text{ROOTED}(R)$.

The following well-known result describes a key property of the graph resulting from an application of this subroutine, see [17]; we include a proof for the sake of completeness.

---
[5]Finding a Hamiltonian cycle in a graph $\hat{G} = (V, \hat{E})$ reduces to this for $k = 2$, by setting $G = (V, \emptyset)$, $c_e = 1$ if $e \in \hat{E}$ and $c_e = \infty$ if $e \in \binom{V}{2} - \hat{E}$.

**Proposition 3.1.** *Let $R \subseteq V$ be a subset of nodes with $|R| = k$, and let the subroutine $\text{ROOTED}(R)$ return the edge set $F'$. Let $(U_0, U_1)$ be a deficient set-pair in $G+F'$. Then $(U_0, U_1)$ is also a deficient set-pair in $G$. Moreover, $R \cap U_0 \neq \emptyset$ and $R \cap U_1 \neq \emptyset$.*

*Proof.* The first part is obvious. Consider the second part. For a contradiction, assume that there is a deficient set-pair $(U_0, U_1)$ in $G + F'$ with $U_0 \cap R = \emptyset$. Pick a node $v \in U_0$. The $k$ internally disjoint paths from $v$ to $\hat{r}$ in the (rooted) graph $\hat{G} + F'$ give $k$ internally disjoint paths from $v$ to the $k$ terminals in $G + F'$. Consider the first node on each path not in $U_0$. Each of these $k$ distinct nodes is in $V - (U_0 \cup U_1)$ because $U_0 \cap R = \emptyset$, by assumption, and there are no edges between $U_0$ and $U_1$, by the definition of set-pair. This gives $p(U_0, U_1) = \max\{0, k - |V - (U_0 \cup U_1)|\} = 0$, a contradiction to the deficiency of the set-pair. □

We apply the following simple corollary to obtain a rogue-free graph.

**Corollary 3.2.** *Let $G = (V, E)$ be a graph, let $R_0$ be a set of $k$ arbitrary nodes of $G$, and let $G'$ be the graph obtained by applying the subroutine $\text{ROOTED}(R_0)$ to $G$. Let $R_1$ be a set of $k$ nodes that is disjoint from every rogue set of $G'$. Then an application of the subroutine $\text{ROOTED}(R_1)$ to $G'$ results in a rogue-free graph.*

### 3.2 Bounding the union of the rogue sets

In this section, we focus on a graph that has been pre-processed by one application of the subroutine $\text{ROOTED}(R)$. We prove Theorem 2.4, namely, the union of all rogue sets is of size $\leq |R|k^2(k-1)$, assuming that every rogue set has a nonempty intersection with $R$. We first need some elementary properties of the function $\gamma(.)$.

**Fact 3.3.** *For all $U, W \subseteq V$, we have*

$$\gamma(U) + \gamma(W) \geq \gamma(U \cap W) + \gamma(U \cup W) \text{ and}$$
$$\gamma(U) + \gamma(W) \geq \gamma(U^* \cap W) + \gamma(U \cap W^*).$$

**Lemma 3.4.** *Let $w_1, w_2$ be two nodes. Let $W_1$ and $W_2$ be inclusion-wise minimal deficient sets such that $w_1 \in W_1 - W_2$, and $w_2 \in W_2 - W_1$ (in other words, for $i \in \{1, 2\}$ and any proper subset of $W_i$, either the subset is not deficient, or the subset does not contain $w_i$). Suppose that $W_1 \cap W_2$ is nonempty. Then, either $w_1 \in \Gamma(W_2)$ or $w_2 \in \Gamma(W_1)$.*

*Proof.* We argue by contradiction. Suppose that $w_1 \notin \Gamma(W_2)$; then $w_1 \in W_2^*$. Similarly, if $w_2 \notin \Gamma(W_1)$, then $w_2 \in W_1^*$. Thus, $w_1 \in W_1 \cap W_2^*$, and $w_2 \in W_2 \cap W_1^*$. We apply the submodularity of $\gamma(.)$ to get
$$2(k-1) \geq \gamma(W_1) + \gamma(W_2) \geq \gamma(W_1 \cap W_2^*) + \gamma(W_2 \cap W_1^*).$$
But, $W_1 \cap W_2^*$ is a proper subset of $W_1$ that contains $w_1$ (it is a proper subset because $W_1 \cap W_2$ is nonempty), hence, by the inclusion-minimal choice of $W_1$, we must have $\gamma(W_1 \cap W_2^*) \geq k$. Similarly, we must have $\gamma(W_2 \cap W_1^*) \geq k$. This gives a contradiction. □

We are now ready to prove Theorem 2.4. For a positive integer $\ell$ we denote the set of integers $\{1, 2, \ldots, \ell\}$ by $[\ell]$.

*Proof of Theorem 2.4.* Let $U_1, U_2, \ldots, U_\ell$ be a smallest family of rogue sets whose union contains every rogue set.

Since $\ell$ is minimum, for each $i \in [\ell]$, the set $U_i$ must contain a "witness node" $w_i$ that is not in any set $U_j, j \neq i$; in other words, $U_i - \bigcup \{U_j \mid j \in [\ell] - \{i\}\}$ is nonempty and we take $w_i$ to be any node of this set.

Next, for each set $U_i$, $i \in [\ell]$, we define $W_i$ to be an inclusion-wise minimal deficient subset of $U_i$ that contains $w_i$. Thus, no proper subset of $W_i$ may contain $w_i$ and be deficient at the same time; the existence of $W_i$ is guaranteed since $U_i$ satisfies both requirements. Let $\mathcal{W}$ denote the family of sets $W_i$: thus, $\mathcal{W} = \{W_1, \ldots, W_\ell\}$.

Each set $W_i$ is also a rogue set, so it must contain a node of $R$ by the condition of the theorem. Consider a fixed but arbitrary node $r \in R$, and focus on all the sets $W_i \in \mathcal{W}$ that contain $r$; let us denote their family by $\mathcal{W}(r) = \{W_i \mid i \in [\ell] \text{ and } r \in W_i\}$. Below, we show that $|\mathcal{W}(r)| \leq k^2$. The same upper bound applies for each node in $R$, yielding $|\mathcal{W}| \leq \sum_{r \in R} |\mathcal{W}(r)| \leq |R|k^2$.

We bound the size of $\mathcal{W}(r)$ by constructing a sequence of sets such that for each set $W_i \in \mathcal{W}(r)$, either $W_i$ is in the sequence, or else $w_i$ (the "witness node" of $W_i$) is in the neighborhood of some set in the sequence. More formally, consider a sequence of sets from $\mathcal{W}(r)$, that is obtained as follows: we start with $\alpha_1$ as the smallest index $i$ such that $W_i \in \mathcal{W}(r)$; assume that the sets $W_{\alpha_1}, \ldots, W_{\alpha_j}$ have been defined; we choose $\alpha_{j+1}$ to be the smallest index $i$ such that $W_i \in \mathcal{W}(r)$ and $w_i \notin \Gamma(W_{\alpha_1}) \cup \Gamma(W_{\alpha_2}) \cup \cdots \cup \Gamma(W_{\alpha_j}) \cup \{w_{\alpha_1}, w_{\alpha_2}, \ldots, w_{\alpha_j}\}$; we stop if there is no such index $i$. Let $\hat{\ell}(r)$ denote the length of this sequence of sets; the last set in the sequence is $W_{\alpha_{\hat{\ell}(r)}}$.

**Claim 3.5.** $\hat{\ell}(r) \leq k$.

*Proof.* Within this proof, let $W = W_{\alpha_{\hat{\ell}(r)}}$. Pick an arbitrary $i \in [\hat{\ell}(r) - 1]$, and apply Lemma 3.4 to the sets $W_{\alpha_i}$ and $W$. Their intersection is nonempty as it contains $r$. Clearly, $w_{\alpha_{\hat{\ell}(r)}}$ (the "witness node" of $W$) is not in $\Gamma(W_{\alpha_i})$, according to the choice of the sets in the sequence. Then, by Lemma 3.4, we have $w_{\alpha_i} \in \Gamma(W)$, and we have $|\Gamma(W)| \leq k - 1$. The conclusion follows: the total number of "witness nodes" of the sets in the sequence is $\leq k$. □

Finally, observe that for each set $W_j \in \mathcal{W}(r)$ that is *not* in the sequence, we have $w_j \in \Gamma(W_{\alpha_1}) \cup \cdots \cup \Gamma(W_{\alpha_{\hat{\ell}(r)}})$. It follows that $|\mathcal{W}(r)| \leq \hat{\ell}(r) + \hat{\ell}(r) \cdot (k-1) \leq k^2$.

Applying the same upper bound for each node $r \in R$, we have $\ell \leq |R|k^2$. It follows that $\bigcup_{i \in [\ell]} U_i$ has size $\leq |R|k^2(k-1)$, since each set $U_i$ has size $\leq k - 1$. □

The proof of the previous theorem relies on two properties, namely, every rogue set is a deficient set, and every rogue set contains a node of the terminal set $R$; but, the bound on the size of rogue sets is used only once, at the end. There is an immediate extension to deficient sets of $G$ of size $\leq s$.

**Theorem 3.6.** *Assume that there exists a set $R \subseteq V$ such that every deficient set of size $\leq s$ has a nonempty intersection with $R$. Then the union of all deficient sets of size $\leq s$ has size $\leq |R|k^2 s$.*

# 4 Iterative rounding in independence-free graphs

Consider two set-pairs $\mathbb{U} = (U_0, U_1)$ and $\mathbb{W} = (W_0, W_1)$. We call $\mathbb{U}$ and $\mathbb{W}$ *nested*, if for some $i,j \in \{0,1\}$, $U_i \supseteq W_{1-j}$ and $W_j \supseteq U_{1-i}$; we call $U_i$ the *dominant piece* of $\mathbb{U}$ with respect to $\mathbb{W}$, and we call $W_j$ the dominant piece of $\mathbb{W}$ w.r.t. $\mathbb{U}$. Note that two nested set-pairs $\mathbb{U}$ and $\mathbb{W}$ are always non-independent, since $uw \in \binom{V}{2}$ covers both set-pairs for arbitrary $u$ and $w$ in the non-dominant pieces of $\mathbb{U}$ and $\mathbb{W}$, respectively. The set-pairs $\mathbb{U}$ and $\mathbb{W}$ are called *crossing* if they are neither independent nor nested. These notions are illustrated in Figure 2. A family $\mathcal{L}$ of set-pairs is called *cross-free* if it has no two crossing members. Note that in an independence-free graph, any two deficient set-pairs in a cross-free family must be nested.

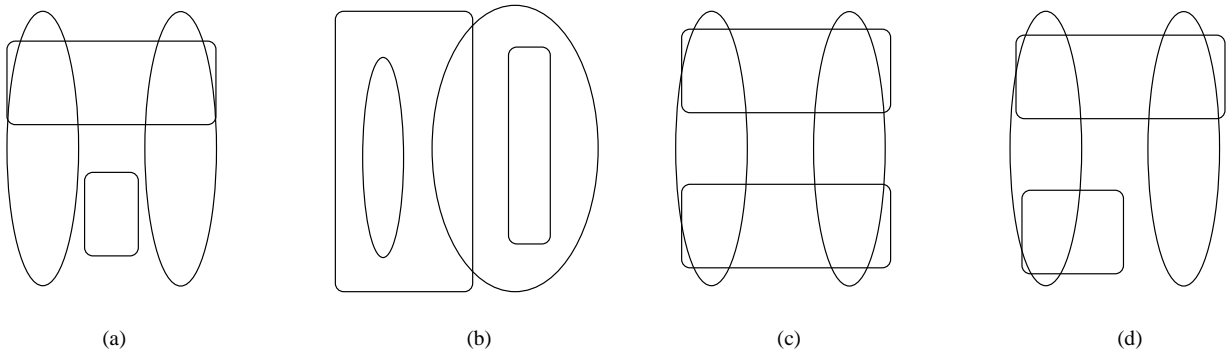

(a)      (b)      (c)      (d)

Figure 2: Relations of set-pairs: *(a)* independent; *(b)* nested; *(c)* crossing with two ways to uncross; *(d)* crossing with only one way to uncross.

Let us now define the uncrossing of set-pairs that cross. This somewhat differs from the standard notion (e.g. [8]) because our set-pairs are unordered. A node $u$ is called a *meeting point* of the set-pairs $\mathbb{U} = (U_0, U_1)$ and $\mathbb{W} = (W_0, W_1)$ if there exists another node $w$ such that $uw \in \binom{V}{2}$ covers both $\mathbb{U}$ and $\mathbb{W}$. Note that two set-pairs have a meeting point if and only if they are non-independent. For any given meeting point $u$, we define two new set-pairs $\mathbb{U} \otimes_u \mathbb{W}$ and $\mathbb{U} \oplus_u \mathbb{W}$ as follows. Let us choose $i,j \in \{0,1\}$ such that the meeting point $u$ lies in $U_i \cap W_j$. Then we define the set-pairs

$$\mathbb{U} \otimes_u \mathbb{W} := (U_i \cup W_j, U_{1-i} \cap W_{1-j}) \text{ and}$$
$$\mathbb{U} \oplus_u \mathbb{W} := (U_i \cap W_j, U_{1-i} \cup W_{1-j}).$$

There is a pair of set-pairs associated with any meeting point $u$, namely, $\mathbb{U} \otimes_u \mathbb{W}$ and $\mathbb{U} \oplus_u \mathbb{W}$; there are at most two such pairs of set-pairs over all possible meeting points (suppose we get one pair for a meeting point in $U_i \cap W_j$ for fixed $i,j \in \{0,1\}$; then we could get another pair for a meeting point in $U_i \cap W_{1-j}$). Figure 2(c) shows two set-pairs that can be uncrossed in two different ways, whereas the set-pairs in Figure 2(d) have a unique way of uncrossing.

If $\mathbb{U}$ and $\mathbb{W}$ are nested set-pairs, then a node $u$ is a meeting point if and only if $u$ belongs to one of the two non-dominant pieces; moreover, for every meeting point $u$, $\{\mathbb{U} \otimes_u \mathbb{W}, \mathbb{U} \oplus_u \mathbb{W}\} = \{\mathbb{U}, \mathbb{W}\}$.

Recall that $\mathcal{S}$ denotes the family of all set-pairs in a graph $G$. A real-valued function $f$ on $\mathcal{S}$ is called *bisubmodular* if for any two set-pairs $\mathbb{U}$ and $\mathbb{W}$ and any meeting point $u$, we have

$$f(\mathbb{U}) + f(\mathbb{W}) \geq f(\mathbb{U} \otimes_u \mathbb{W}) + f(\mathbb{U} \oplus_u \mathbb{W}).$$

For any nonnegative vector $\boldsymbol{x} : E \to \mathbb{R}_+$ on the edges, the corresponding function on set-pairs, $x(\delta(W)) = \sum_{e \in \delta(W)} x_e$, is bisubmodular.

A non-negative, integer-valued function $f$ on $\mathcal{S}$ is called *positively crossing bisupermodular* if for any two crossing set-pairs $\mathbb{U}$ and $\mathbb{W}$ with $f(\mathbb{U}) > 0$ and $f(\mathbb{W}) > 0$ and a meeting point $u$, we have

$$f(\mathbb{U}) + f(\mathbb{W}) \leq f(\mathbb{U} \otimes_u \mathbb{W}) + f(\mathbb{U} \oplus_u \mathbb{W}).$$

**Claim 4.1.** *The deficiency function $p : \mathcal{S} \to \mathbb{R}_+$ defined by (1) is positively crossing bisupermodular.*

*Proof.* Consider two crossing set-pairs $\mathbb{U}$ and $\mathbb{W}$ with $p(\mathbb{U}) > 0$, $p(\mathbb{W}) > 0$, and let $u$ be a meeting point. We have to show that

$$(k - |\Gamma(\mathbb{U})|) + (k - |\Gamma(\mathbb{W})|) \leq$$
$$\max\{0, k - |\Gamma(\mathbb{U} \otimes_u \mathbb{W})|\} + \max\{0, k - |\Gamma(\mathbb{U} \oplus_u \mathbb{W})|\}.$$

This holds because $|\Gamma(\mathbb{U})| + |\Gamma(\mathbb{W})| = |\Gamma(\mathbb{U} \otimes_u \mathbb{W})| + |\Gamma(\mathbb{U} \oplus_u \mathbb{W})|$, or equivalently, $|U_0 \cup U_1| + |W_0 \cup W_1|$ is equal to the sum of the size of the union of the two pieces of $\mathbb{U} \otimes_u \mathbb{W}$ and the size of the union of the two pieces of $\mathbb{U} \oplus_u \mathbb{W}$. □

As in [8], a basic optimal solution to (LP-VC) can be found in polynomial time using the ellipsoid algorithm (see [12, Theorem 6.4.9]). The next result characterizes a basic solution of (LP-VC) via a cross-free family of set-pairs. The theorem holds for arbitrary input graphs, without assuming independence-freeness. Analogous results are well-known in the iterative rounding literature; the proof follows the standard lines (e.g. [21, Theorem 4.1.5], [3, Theorem 3.3]). We defer the proof to the appendix.

**Theorem 4.2.** *Let $\boldsymbol{x}$ be a basic solution of (LP-VC) such that $x_e < 1$ for all edges $e \in \binom{V}{2}$. Let $\mathrm{supp}(x)$ denote the support of $\boldsymbol{x}$, that is, the set of edges $e \in \binom{V}{2}$ with $x_e > 0$, and for a set-pair $\mathbb{U}$, let $\chi(\mathbb{U}) = \delta(\mathbb{U}) \cap \mathrm{supp}(x)$ denote the set of edges in $\mathrm{supp}(x)$ covering $\mathbb{U}$. Then there exists a cross-free family $\mathcal{L}$ of deficient set-pairs such that:*

(i) $|\mathcal{L}| = |\mathrm{supp}(x)|$.

(ii) *The vectors $\chi(\mathbb{U})$, $\mathbb{U} \in \mathcal{L}$, are linearly independent.*

*The same hold if $p$ in (LP-VC) is replaced by an arbitrary positively crossing bisupermodular function on set-pairs.*

In what follows, we prove Theorem 2.1, based on the "fractional token" technique of Nagarajan et al. [24]. The proof straightforwardly extends to the more general setting where $p$ is an arbitrary positively crossing bisupermodular function; by independence-freeness we mean that there are no two set-pairs with positive $p$ values that are independent.

We first state some properties of cross-free families of set-pairs in an independence-free graph. For a set-pair $\mathbb{U} = (U_0, U_1)$ and $i \in \{0, 1\}$, let us call $U_i$ the *tail* of $\mathbb{U}$ if $|U_i| < |U_{1-i}|$; moreover, if $|U_0| = |U_1|$, let us arbitrarily designate one of the pieces to be the tail. The piece different from the tail is called the *head*. We denote the tail and head of a set-pair $\mathbb{U}$ by $\mathbb{U}_t$ and $\mathbb{U}_h$, respectively. The next lemma will be applied for the cross-free family $\mathcal{L}$ as in Theorem 4.2. As usual, we say that a family of sets $\mathcal{S}$ is *laminar* if for any $X, Y \in \mathcal{S}$, either $X \subseteq Y$ or $Y \subseteq X$ or $X \cap Y = \emptyset$.

**Lemma 4.3.** *Suppose that $G = (V, E)$ is an independence-free graph. Let $\mathcal{L}$ be a cross-free family of set-pairs.*

(i) *The tails of the set-pairs in $\mathcal{L}$ form a laminar family $\{\mathbb{U}_t \mid \mathbb{U} \in \mathcal{L}\}$ that we denote by $\mathcal{L}_t$. Suppose that we have two disjoint tails in $\mathcal{L}_t$; then each tail is a subset of the other set-pair's head.*

(ii) *Suppose that an edge $e = uw$ has exactly one endnode in a tail $\mathbb{U}_t \in \mathcal{L}_t$, say $u \in \mathbb{U}_t$, and suppose that $e$ does not cover the set-pair $\mathbb{U}$. Then, if there is a tail $\mathbb{Y}_t \in \mathcal{L}_t$ that contains $w$ (the other endnode of $e$), then $\mathbb{U}_t \subseteq \mathbb{Y}_t$.*

*Proof.* **(i)** Suppose that the tails of two set-pairs $\mathbb{U}, \mathbb{W} \in \mathcal{L}$ intersect properly (that is, $\mathbb{U}_t \cap \mathbb{W}_t, \mathbb{U}_t - \mathbb{W}_t, \mathbb{W}_t - \mathbb{U}_t$ are all nonempty). By cross-freeness, $\mathbb{U}$ and $\mathbb{W}$ are either nested or independent, but the latter is excluded because the graph is assumed to be independence-free. Hence $\mathbb{U}$ and $\mathbb{W}$ are nested with $\mathbb{U}_t$ and $\mathbb{W}_t$ being the dominant pieces, and consequently, $\mathbb{U}_h \subseteq \mathbb{W}_t$ and $\mathbb{W}_h \subseteq \mathbb{U}_t$. This implies
$$|\mathbb{U}_h| \leq |\mathbb{W}_t| \leq |\mathbb{W}_h| \leq |\mathbb{U}_t| \leq |\mathbb{U}_h|.$$
Equality must hold throughout. Therefore $\mathbb{U}$ and $\mathbb{W}$ are identical set-pairs, a contradiction. Hence, the tails of the set-pairs in $\mathcal{L}$ form a laminar family. The second part of the claim also follows.
**(ii)** Suppose that $w \in \mathbb{Y}_t$. Since $\mathbb{Y}_t$ and $\mathbb{U}_t$ belong to a laminar family, either the two tails are disjoint or $\mathbb{Y}_t$ is a superset of $\mathbb{U}_t$ (it cannot be a subset because $w \in \mathbb{Y}_t$ and $w \notin \mathbb{U}_t$).

Suppose that $\mathbb{Y}_t$ and $\mathbb{U}_t$ are disjoint. By part (i), we have $w \in \mathbb{Y}_t \subseteq \mathbb{U}_h$. Then the edge $e = uw$ would cover the set-pair $\mathbb{U}$; this contradicts the statement of (ii). $\square$

*Proof of Theorem 2.1.* By way of contradiction, suppose that $\boldsymbol{x}$ is a basic feasible solution with $x_e < 1/2$ for all edges $e \in \text{supp}(x)$.

By Theorem 4.2, $\boldsymbol{x}$ is associated with a cross-free family of set-pairs, call it $\mathcal{L}$; also, let $\mathcal{L}_t$ be the laminar family of tails. We define parent, child, smallest set containing a specified node, etc. in the usual way for the laminar family of tails $\mathcal{L}_t$. Moreover, for ease of notation, we use the same terms for the set-pairs in $\mathcal{L}$, e.g., if $\mathbb{U}_t$ has two children $\mathbb{W}_{i:t}$ and $\mathbb{W}_{j:t}$ in $\mathcal{L}_t$, then we say that $\mathbb{U}$ has two children $\mathbb{W}_i$ and $\mathbb{W}_j$ in $\mathcal{L}$.

We will show that $|\text{supp}(x)| > |\mathcal{L}_t|$, thus contradicting Theorem 4.2. (Note that $|\mathcal{L}_t| = |\mathcal{L}|$.)

We assign a unit token to each edge in $\text{supp}(x)$, and then we redistribute tokens to the sets in $\mathcal{L}_t \cup \{V\}$ in such a way that every set in $\mathcal{L}_t$ gets at least one unit of token, and $V$ also gets some positive amount. This will imply $|\text{supp}(x)| > |\mathcal{L}_t|$.

Consider any edge $e = uw \in \text{supp}(x)$. Let $\mathbb{U}_t \in \mathcal{L}_t \cup \{V\}$ be the smallest tail that contains the endnode $u$ of $e$, and let $\mathbb{W}_t \in \mathcal{L}_t \cup \{V\}$ be the smallest tail that contains the endnode $w$ of $e$. The unit token of $e$ is redistributed to the tails in $\mathcal{L}_t$ using the following rules.

(i) Suppose that $\mathbb{U}_t$ and $\mathbb{W}_t$ are disjoint; then we assign $x_e$ tokens to each of $\mathbb{U}_t$ and $\mathbb{W}_t$, and we assign $1 - 2x_e$ tokens to the smallest tail in $\mathcal{L}_t \cup \{V\}$ that contains both $u$ and $w$.

(ii) Otherwise, one of the tails $\mathbb{U}_t$ or $\mathbb{W}_t$ is a subset of the other one; w.l.o.g. suppose that $\mathbb{W}_t \subseteq \mathbb{U}_t$; then we assign $x_e$ tokens to $\mathbb{W}_t$, and we assign $1 - x_e$ tokens to the smallest tail $\mathbb{Z}_t \in \mathcal{L}_t \cup \{V\}$ such that $u, w \in \mathbb{Z} \cup \Gamma(\mathbb{Z})$.

Observe that two cases could arise within rule (ii): we have $w \in \mathbb{Z}_t$ or $w \in \Gamma(\mathbb{Z})$. In the first case, $\mathbb{Z} = \mathbb{U}$ follows, while in the second case, $\mathbb{Z}_t \subsetneq \mathbb{U}_t$ is possible.

We claim that each tail in $\mathcal{L}_t$ gets at least one token. Consider a set-pair $\mathbb{U} \in \mathcal{L}$, and let it have $q$ children $\mathbb{W}_1, \mathbb{W}_2, \ldots, \mathbb{W}_q$ (possibly, $q = 0$). We now focus on the set of edges given by the symmetric difference of $\chi(\mathbb{U})$ and $\bigcup_{i=1}^{q} \chi(\mathbb{W}_i)$ (recall $\chi(\mathbb{U}) = \delta(U) \cap \text{supp}(x)$), and we partition this set into three sets $A, B, C$ as follows:

- $A$ is the set of edges with (exactly) one endnode in $\mathbb{U}_t - \bigcup_{i=1}^{q} \mathbb{W}_{i:t}$ and that cover $\mathbb{U}$;

- $B$ is the set of edges with both endnodes in $\bigcup_{i=1}^{q} \mathbb{W}_{i:t}$ and covering two of the children $\mathbb{W}_1, \mathbb{W}_2, \ldots, \mathbb{W}_q$;

- $C$ is the set of edges that cover one of the children $\mathbb{W}_1, \mathbb{W}_2, \ldots, \mathbb{W}_q$, but that do not cover $\mathbb{U}$.

Note that if an edge has an endnode in $\mathbb{W}_{i:t}$ and covers $\mathbb{U}$, then it must also cover $\mathbb{W}_i$. By subtracting the equations of the children from the equation of $\mathbb{U}$, we get

$$p(\mathbb{U}) - \sum_{i=1}^{q} p(\mathbb{W}_i) = x(\delta(\mathbb{U})) - \sum_{i=1}^{q} x(\delta(\mathbb{W}_i)) =$$
$$= x(A) - 2x(B) - x(C). \tag{2}$$

**Claim 4.4.** *If $e \in A$ then $x_e$ tokens from $e$ are assigned to $\mathbb{U}$. If $e \in B$ then $1 - 2x_e$ tokens from $e$ are assigned to $\mathbb{U}$. Finally, if $e \in C$, then $1 - x_e$ tokens from $e$ are assigned to $\mathbb{U}$.*

*Proof.* The first two claims are straigtforward by the definitions. Consider any edge $e = uw \in C$. Let $\mathbb{W}_i$ be the child covered by $e$ with $u \in \mathbb{W}_{i:t}$. We claim that the token of $e$ is distributed according to rule (ii) and $1 - x_e$ is allocated to $\mathbb{U}$. This clearly holds if $w \in \mathbb{U}_t$ as $\mathbb{U}_t$ is the smallest tail in $\mathcal{L}_t$ containing $w$; also, the tail $\mathbb{Z}_t$ of rule (ii) is equal to $\mathbb{U}_t$.

Next, assume $w \notin \mathbb{U}_t$. Then $w \in \Gamma(\mathbb{U})$ since $uw$ does not cover $\mathbb{U}$. Let $\mathbb{Y}_t \in \mathcal{L}_t$ be the smallest tail containg $w$. Then Lemma 4.3(ii) is applicable for $\mathbb{U}$ and $\mathbb{Y}$, yielding $\mathbb{U}_t \subseteq \mathbb{Y}_t$. Consequently, the smallest tails containing $u$ and $w$ cannot be disjoint, and therefore rule (ii) applies. Since $\mathbb{W}_i$ is a child of $\mathbb{U}$ and $uw$ covers $\mathbb{W}_i$, it also follows that $\mathbb{U}$ is the set-pair with the smallest tail containing $u$ but not covered by $e$. Consequently, $\mathbb{U}$ receives $1 - x_e$ tokens from $e$. □

$A \cup B \cup C$ must be nonempty; otherwise, we have $\chi(\mathbb{U}) = \sum_{i=1}^{q} \chi(\mathbb{W}_i)$, contradicting linear independence. Using the above Claim and (2), we obtain the following lower bound on the amount of tokens received by $\mathbb{U}$ (it may get even more).

$$\sum_{e \in A} x_e + \sum_{e \in B}(1 - 2x_e) + \sum_{e \in C}(1 - x_e) =$$
$$x(A) + (|B| - 2x(B)) + (|C| - x(C)) =$$
$$|B| + |C| + x(A) - 2x(B) - x(C) =$$
$$|B| + |C| + p(\mathbb{U}) - \sum_{i=1}^{q} p(\mathbb{W}_i)$$

Since $A \cup B \cup C$ is nonempty, and $0 < x_e < \frac{1}{2}$ for each edge $e$, the above quantity is strictly positive (by the LHS expression). On the other hand, it is integer (by the RHS expression). Hence, $\mathbb{U}$ gets at least one token.

Finally, we derive the contradiction by showing that $V$ received a positive amount of tokens. Consider any (inclusion-wise) maximal tail $\mathbb{U}_t \in \mathcal{L}_t$; there must be at least one edge $f = vw$ that covers $\mathbb{U}$. Then $V$ receives either the $1 - 2x_f$ tokens assigned by rule (i) for $f$ or the $1 - x_f$ tokens assigned by rule (ii) for $f$. This completes the proof of Theorem 2.1. □

## 5 Algorithmic aspects

Our algorithm starts by applying the subroutine ROOTED($R_0$), for an arbitrary subset $R_0 \subseteq V$ of size $k$. Let $G_0$ denote the resulting graph; thus, $G_0$ contains all of the edges added by ROOTED($R_0$). By Corollary 3.2 and Theorem 2.4, if $n \geq k^3(k-1)+k$, then there exists a set of nodes $R_1$, $|R_1| = k$ disjoint from every rogue set of $G_0$, and the application of subroutine ROOTED($R_1$) results in a rogue-free graph $G_1$. Clearly, $G_1$ is also independence-free (by Fact 2.3). Hence, by Theorem 2.1, iterative rounding can be applied to find an augmenting edge set of cost $\leq 2\mathrm{opt}(G_1) \leq 2\mathrm{opt}(G_0) \leq 2\mathrm{opt}(G)$.

Whereas the existence of an appropriate set $R_1$ is guaranteed if $n \geq k^3(k-1)+k$, it is a nontrivial algorithmic task to find one. If $k^3(k-1) + k \leq n < k^4(k-1) + k$, then we apply a brute-force method described in Section 5.1 that is based on a stronger version of Theorem 2.1. This method works for larger values of $n$ as well, but in Section 5.2, we present a different and more efficient algorithm that is based on submodular function minimization for the case of $n \geq k^4(k-1) + k$.

### 5.1 Small values of $n$

In this part, we assume that $k^3(k-1) + k \leq n < k^4(k-1) + k$. Our method is based on the following strengthening of Theorem 2.1 that allows the input graph to contain deficient set-pairs that are independent.

**Theorem 5.1.** *Let $G = (V, E)$ be an arbitrary graph, and let $x$ be a basic feasible solution to (LP-VC). Then either there exists an edge $e$ with $x_e \geq 1/2$, or we can find a rogue set efficiently.*

*Proof.* The key point is to show that a rogue set can be found efficiently, if $x_e < 1/2$ for each edge $e$, where $x$ is a basic feasible solution of (LP-VC). This is based on the following claim.

**Claim 5.2.** *If $x_e < 1/2$ for each edge $e$, then there exist two independent deficient set-pairs $\mathbb{U}$ and $\mathbb{W}$ with $p(\mathbb{U}) = x(\delta(\mathbb{U}))$, $p(\mathbb{W}) = x(\delta(\mathbb{W}))$.*

Assuming this claim, let us add every $e \in \binom{V}{2}$ as a fractional edge of value $x_e$ to $G$. The resulting (fractional) graph is $k$-connected, and its minimum node cuts correspond to tight set-pairs (set-pairs satisfying $x(\delta(\mathbb{W})) = p(\mathbb{W})$).

Using standard network-flow techniques (bidirect every edge and replace every node by a capacitated directed edge) we can compute a minimum node cut separating any two nodes $u, w \in V$ by a max-flow min-cut computation. Moreover, the computation also finds the unique inclusionwise-minimal one among the minimum $u, w$ cuts. Let us compute the inclusionwise-minimal minimum $u, w$ cut for every pair $u, w \in V$. In Claim 5.2, at least one piece of $\mathbb{U}$ or $\mathbb{W}$ is a rogue set, and consequently, one of these inclusionwise-minimal sets found by network-flow techniques must be a rogue set.

It is left to prove Claim 5.2. Consider the cross-free family $\mathcal{L}$ as in Theorem 4.2; note that independence-freeness is not assumed. If this family is independence-free, then the entire argument in the proof of Theorem 2.1 carries over, showing that there exists an edge $e$ with $x_e \geq \frac{1}{2}$, in contradiction to our assumption. Consequently, $\mathcal{L}$ must contain two independent set-pairs, verifying the claim. □

The algorithm starts by applying ROOTED($R_0$) for an arbitrary set $R_0 \subseteq V$ of size $k$, to obtain the graph $G_0$. Then we repeat the following two steps: *(i)* apply the subroutine ROOTED($R_1$) in $G_0$ (with different sets of nodes $R_1 = R_1', R_1'', \ldots$ each of size $k$), and *(ii)* apply iterative rounding in the resulting graph.

In more detail, let $S \subseteq V$ denote the "forbidden set" for the node set $R_1$; we initialize $S = R_0$. In every outer loop of the algorithm, we pick a set of nodes $R_1$ of size $k$ disjoint from $S$ and run the subroutine ROOTED($R_1$) in $G_0$, followed by a run of the iterative rounding algorithm; we proceed with the latter as long as there exists an edge $e$ with $x_e \geq 1/2$. If we successfully arrive at a $k$-connected graph by this method, then the overall algorithm terminates. If we fail to obtain a $k$-connected graph, then we can find a rogue set $X$ by Theorem 5.1. Clearly, $X$ was already a rogue set in the graph $G_0$. Let us replace $S$ by $S \cup X$, and restart in $G_0$ with a new set $R_1$ disjoint from $S$. Note that the size of $S$ increases by at least one since $R_1 \cap S = \emptyset$, and $R_1 \cap X \neq \emptyset$ by Proposition 3.1. Since the union of all rogue sets in $G_0$ has size $\leq k^3(k-1)$, the entire procedure may be repeated at most $k^3(k-1) - k$ times.

## 5.2 Large values of $n$

In this part, we focus on the case $k^4(k-1) + k \leq n$. Our plan is to identify a set $B \subseteq V$ such that $|B| \leq k^4(k-1)$ and $B$ contains every rogue set. After that, we can easily find an appropriate set of $k$ terminals $R_1$ that is disjoint from $B$.

Let us define the function $h : 2^V \to \mathbb{R}_+$ by $h(X) = |X| + (k-1)\gamma(X)$. The following claim is straightforward.

**Claim 5.3.** (i) *For every rogue set $X$, $h(X) \leq k(k-1)$.*
(ii) *If $h(X) \leq k(k-1)$ for a set $\emptyset \neq X \subseteq V$, then $X$ is a deficient set and $|X| \leq k(k-1)$.*

We define $B$ to be the union of all sets $X$ with $h(X) \leq k(k-1)$. By part (i) of the claim, $B$ contains all rogue sets. By part (ii) and Theorem 3.6, we get $|B| \leq k^4(k-1)$.

To find $B$, observe that $h$ is a fully submodular function. Consequently, for every $v \in V$, we can find the minimal value of $h(X)$ over all sets $X$ containing $v$ in strongly polynomial time, see [27, 13]. These algorithms can also be used to find the unique largest set $X$ containing $v$ that achieves the above minimum value of $h(.)$.

The subroutine for finding $B$ proceeds as follows. We start with $A, B = \emptyset$. In each step, we take a node $v \in V - (A \cup B)$, and apply the subroutine for submodular function minimization. If the minimum value is greater than $k(k-1)$, then we add $v$ to the set $A$. Otherwise, let $X$ be the minimizer set that has the largest size. Replace $B$ by $B \cup X$ and proceed to the next node in $V - (A \cup B)$. The subroutine terminates once $A \cup B = V$ is attained.

Hence the algorithm for minimum cost $k$-connectivity augmentation first performs ROOTED($R_0$) for an arbitrary subset $R_0 \subseteq V$ of size $k$, resulting in $G_0$. Then we apply the above subroutine for finding the set $B$ in $G_0$, and then we choose an arbitrary $R_1 \subseteq V - B$, $|R_1| = k$, and perform ROOTED($R_1$). Finally, we apply iterative rounding in the resulting independence-free graph.

# 6   Discussion

In this paper, we only cover the assymptotic setting of $k$-connectivity augmentation, for the case $n \geq k^3(k-1) + k$, leaving the case of all values of $n$ open. An immediate way to improve the result is to replace the bound $k^3(k-1)$ on the union of rogue sets in Theorem 2.4 by a smaller function of $k$. By the time of the submission, this has already been improved by Nutov [10], giving a simple proof of the stronger bound $(k-1)^3 - k$.

Also, note that the first set of terminals is chosen arbitrarly; further improvement might be possible by a clever choice. Yet it seems difficult to obtain an $O(1)$ approximation guarantee for all values of $n$ using these tools only, and substantial new insights may be needed to resolve this, e.g., as in [19, 4, 25], as compared to [3]. Note that if $n < 2k$, then our method is entirely void: making a graph rogue-free is equivalent to the original connectivity augmentation problem.

An important special case of our problem is the min-cost augmentation-by-one problem, i.e., when the input graph is already $(k-1)$-connected. The paper [3] gave a 6-approximation for the asymptotic setting by applying the Frank-Tardos algorithm 3 times based on a result of Mader [23] on 3-critical graphs. Our methods do not seem to give any improvement on 6-approximation for augmentation-by-one in the asymptotic setting, but Nutov [25] gives a 5-approximation.

Our result only concerns undirected graphs and does not apply for directed graphs. This is in contrast with most of the literature (see [19, 4, 25]), where the undirected problem is essentially solved via a reduction to the more general setting of directed graphs. However, it seems that undirected set-pairs have certain advantageous properties not shared by their directed counterparts. In particular, the right notion of independence-freeness for directed graphs is not clear; forbidding all independence in the directed sense seems too restrictive. A good candidate for the notion of rogue sets could be the sets of size less than $k$ that are both in-deficient and out-deficient. Yet we were not able to prove any analogue of Theorem 2.1 even assuming rogue-free directed graphs in this sense. Also, bounding the size of the union of such rogue sets seems more challenging.

There is a line of research focusing on degree-bounded problems in network design, i.e., finding a min-cost subgraph subject to both connectivity requirements and bounds on the degrees of the nodes. For the degree-bounded (edge-connectivity) SNDP, bicriteria approximations were given by Lau et al. [20] and by Lau and Singh [22]. Recently, Nutov [26] and Fukunaga and Ravi [11] have presented such results for several degree-bounded problems with node-connectivity requirements. In particular, for min-cost degree bounded $k$-node-connected spanning subgraphs [11] gives a $(O(k), 2b(v) + O(k^2))$ bicriteria approximation, i.e., given an upper-bound of $b(v)$ on the degree of each node $v$, [11] finds a solution subgraph of cost $O(k)$ times the optimal cost of the relevant LP relaxation such that the degree of each node $v$ is $\leq 2b(v) + O(k^2)$.

It may be possible to extend our approach to obtain an $(O(1), O(1)b(v))$ bicriteria approximation for the asymptotic setting. Indeed, instead of using the Frank-Tardos algorithm, one may apply the $(4, 2b(v) + O(k))$ bicriteria approximation for degree-bounded directed $k$-outconnectivity from [11]. Lau et al. [20], [21] extended Jain's iterative rounding results and token arguments to give a $(2, 2b(v) + 3)$ bicriteria approximation for the degree-bounded SNDP. It may be possible to extend these results to the setting of positively crossing bisupermodular requirements in independence-free graphs. However, such an extension does not seem straightforward. Combining these results with Theorem 2.4 would give an $(O(1), O(1)b(v))$ bicriteria approximation for the degree-bounded version of our problem in the asymptotic setting.

Our algorithm first applies a combinatorial pre-processing, and then it solves a continuous relaxation (namely, an LP relaxation) and rounds the fractional solution to get an integer solution. Neither method by itself is known to achieve good approximation guarantees (not even polylog in $k$), but the combined method achieves a constant approximation guarantee in the asymptotic setting. An analogous scheme is applied in an entirely different context by Karger, Motwani and Sudan [16] for coloring 3-colorable graphs with $\tilde{O}(n^{1/3})$ colors. A randomized rounding of a semidefinite programming relaxation (SDP) is an efficient tool, however, it performs much better for graphs with low maximum degree. The best approximation guarantee can be obtained by first eliminating the high degree nodes using a combinatorial preprocessing based on Widgerson's [28] algorithm.

## Acknowledgments

We are grateful to a number of colleagues, in particular, Takuro Fukunaga, Bundit Laekhanukit, Zeev Nutov, and R.Ravi, for useful discussions and comments that helped to improve the presentation of the paper. We also thank Santosh Vempala for pointing out the analogy with [16].

## Appendix

### Proof of Theorem 4.2

Given a basic solution $\boldsymbol{x}$ as in Theorem 4.2, let $\mathcal{F} = \{\mathbb{U}: x(\delta(\mathbb{U})) = p(\mathbb{U})\}$ denote the family of tight constraints. It is well-known using basic linear algebra that rank$\{\chi(\mathbb{U}), \mathbb{U} \in \mathcal{F}\} = |\text{supp}(x)|$, see [21, Lemma 2.1.4]. Hence the theorem will be a consequence of the following lemma, by choosing a maximal family $\mathcal{H} \subseteq \mathcal{F}$ with $\chi(\mathbb{U}), \mathbb{U} \in \mathcal{H}$ being cross-free.

**Lemma 6.1.** *Let $\mathcal{H}$ be a maximal cross-free subfamily of deficient sets in $\mathcal{F}$. Then span$\{\chi(\mathbb{U}), \mathbb{U} \in \mathcal{F}\} = $ span$\{\chi(\mathbb{U}), \mathbb{U} \in \mathcal{H}\}$.*

The proof needs the following claim.

**Claim 6.2.** *Assume $\mathbb{U}, \mathbb{W} \in \mathcal{F}$ have a meeting point $u$, and $p(\mathbb{U}), p(\mathbb{W}) > 0$. Then also $\mathbb{U} \otimes_u \mathbb{W}$, $\mathbb{U} \oplus_u \mathbb{W} \in \mathcal{F}$, and*

$$\chi(\mathbb{U}) + \chi(\mathbb{W}) = \chi(\mathbb{U} \otimes_u \mathbb{W}) + \chi(\mathbb{U} \oplus_u \mathbb{W}). \tag{3}$$

*Proof.* Applying the positively crossing bisupermodularity of $p(.)$ (Claim 4.1) and the bisubmodularity of $x(\delta(.))$, we get that

$$x(\delta(\mathbb{U})) + x(\delta(\mathbb{W})) = p(\mathbb{U}) + p(\mathbb{W}) \leq$$
$$p(\mathbb{U} \otimes_u \mathbb{W}) + p(\mathbb{U} \oplus_u \mathbb{W}) \leq x(\delta(\mathbb{U} \otimes_u \mathbb{W})) + x(\delta(\mathbb{U} \oplus_u \mathbb{W})) \leq$$
$$x(\delta(\mathbb{U})) + x(\delta(\mathbb{W})),$$

and hence equality must hold everywhere. This implies both parts of the claim. □

*Proof of Lemma 6.1.* For a contradiction, assume $\mathcal{H}$ is a maximal cross-free subfamily of deficient sets in $\mathcal{F}$, yet span$\{\chi(\mathbb{U}), \mathbb{U} \in \mathcal{H}\} \subsetneq$ span$\{\chi(\mathbb{U}), \mathbb{U} \in \mathcal{F}\}$. For any $\mathbb{W} \in \mathcal{F} - \mathcal{H}$, let cross$(\mathbb{W}, \mathcal{H})$ denote the number of set-pairs in $\mathcal{H}$ crossing $\mathbb{W}$. Let us pick $\mathbb{W}$ such that $\chi(\mathbb{W}) \notin$ span$\{\chi(\mathbb{U}), \mathbb{U} \in \mathcal{F}\}$, and cross$(\mathbb{W}, \mathcal{H})$ is minimal. Clearly, cross$(\mathbb{W}, \mathcal{H}) \geq 1$ as otherwise we could extend $\mathcal{H}$ by $\mathbb{W}$ keeping the cross-free property. Let us choose $\mathbb{U} \in \mathcal{H}$ such that $\mathbb{U}$ and $\mathbb{W}$ cross; let $u$ be a meeting point of $\mathbb{U}$ and $\mathbb{W}$. Clearly, $p(\mathbb{U}) > 0$, as otherwise $x(\delta(\mathbb{U})) = p(\mathbb{U}) = 0$, and hence $\chi(\mathbb{U}) = 0$, contradicting the choice of $\mathbb{U}$.

**Claim 6.3.** *If $p(\mathbb{U} \otimes_u \mathbb{W}) > 0$, then $cross(\mathbb{U} \otimes_u \mathbb{W}, \mathcal{H}) < cross(\mathbb{W}, \mathcal{H})$. If $p(\mathbb{U} \oplus_u \mathbb{W}) > 0$, then $cross(\mathbb{U} \oplus_u \mathbb{W}, \mathcal{H}) < cross(\mathbb{W}, \mathcal{H})$.*

*Proof.* We verify the claim for $\mathbb{U} \otimes_u \mathbb{W}$; the proof is analogous for $\mathbb{U} \oplus_u \mathbb{W}$. Assume $p(\mathbb{U} \otimes_u \mathbb{W}) > 0$, i.e. $\mathbb{U} \otimes_u \mathbb{W}$ is deficient. Observe that whereas $\mathbb{U}$ and $\mathbb{W}$ cross, $\mathbb{U}$ and $\mathbb{U} \otimes_u \mathbb{W}$ are nested. The claim follows by showing that whenever $\mathbb{U} \otimes_u \mathbb{W}$ crosses some $\mathbb{T} \in \mathcal{H}$, then $\mathbb{W}$ and $\mathbb{T}$ also cross.

W.l.o.g., assume $u \in U_0 \cap W_0$, that is, $\mathbb{U} \otimes_u \mathbb{W} = (U_0 \cup W_0, U_1 \cap W_1)$. For a contradiction, assume there exists a $\mathbb{T} \in \mathcal{H}$ such that $\mathbb{U} \otimes_u \mathbb{W}$ and $\mathbb{T}$ cross, but $\mathbb{W}$ and $\mathbb{T}$ are either independent or nested. As $\mathcal{H}$ is cross-free, $\mathbb{U}$ and $\mathbb{T}$ are also either independent or nested.

**Case I.** $\mathbb{T}$ is independent from both $\mathbb{U}$ and $\mathbb{W}$. Clearly, any edge of $\binom{V}{2}$ covering $\mathbb{U} \otimes_u \mathbb{W} = (U_0 \cup W_0, U_1 \cap W_1)$ must cover either $\mathbb{U}$ or $\mathbb{W}$. Consequently, if $\mathbb{U} \otimes_u \mathbb{W}$ and $\mathbb{T}$ are non-independent, then either $\mathbb{U}$ and $\mathbb{T}$ are non-independent or $\mathbb{W}$ and $\mathbb{T}$ are non-independent. This gives a contradiction.

**Case II.** $\mathbb{T}$ is nested with both $\mathbb{U}$ and $\mathbb{W}$. It can be seen that the dominant piece of $\mathbb{T}$ w.r.t. $\mathbb{U}$ is the same as the dominant piece of $\mathbb{T}$ w.r.t. $\mathbb{W}$, and hence it follows that for some $i, j, \ell \in \{0, 1\}$, $T_\ell \supseteq U_i \cup W_j$, $T_{1-\ell} \subseteq U_{1-i} \cap W_{1-j}$. For every possible choice of indices $i, j \in \{0, 1\}$, it can be verified that $\mathbb{U} \otimes_u \mathbb{W}$ and $\mathbb{T}$ are also nested. This gives a contradiction.

**Case III.** $\mathbb{T}$ is independent with either of $\mathbb{U}$ and $\mathbb{W}$ and nested with the other one. By symmetry, we may assume w.l.o.g. that $\mathbb{U}$ and $\mathbb{T}$ are independent, whereas $\mathbb{W}$ and $\mathbb{T}$ are nested. Assume first that $W_0$ is the dominant piece of $\mathbb{W}$ w.r.t. $\mathbb{T}$. Then $\mathbb{U} \otimes_u \mathbb{W}$ and $\mathbb{T}$ are also nested, with $U_0 \cup W_0$ being the dominant part. Next, assume that $W_1$ is the dominant piece of $\mathbb{W}$ w.r.t. $\mathbb{T}$, and let $T_0$ be the dominant piece of $\mathbb{T}$ w.r.t. $\mathbb{W}$; thus, we have $W_0 \subseteq T_0$.

We claim that $\mathbb{U} \otimes_u \mathbb{W}$ and $\mathbb{T}$ must be independent. Indeed, let $pq \in \binom{V}{2}$ be an edge covering both, with $p \in T_0$, $q \in T_1$. We have two cases:

(a) $p \in U_0 \cup W_0$, $q \in U_1 \cap W_1$, or

(b) $q \in U_0 \cup W_0$, $p \in U_1 \cap W_1$.

Both cases contradict the independence of $\mathbb{U}$ and $\mathbb{T}$. In case (a), we have $q \in T_1 \cap U_1 \cap W_1 \subseteq U_1 \cap T_1$, and this is a contradiction because the meeting point $u$ is in $U_0 \cap W_0 \subseteq U_0 \cap T_0$: the edge $uq \in \binom{V}{2}$ covers both $\mathbb{U}$ and $\mathbb{T}$. In case (b), since $q \notin T_0$ and $T_0 \supseteq W_0$, it follows that $q \in U_0 - W_0$, and hence $pq$ covers both $\mathbb{U}$ and $\mathbb{T}$.

This completes the proof of Claim 6.3. □

By Claim 6.2, both $\mathbb{U} \otimes_u \mathbb{W}$ and $\mathbb{U} \oplus_u \mathbb{W}$ are in $\mathcal{F}$, and (3) holds. Let us first show that $p(\mathbb{U} \otimes_u \mathbb{W}), p(\mathbb{U} \oplus_u \mathbb{W}) > 0$. Indeed, if $p(\mathbb{U} \otimes_u \mathbb{W}) = 0$, then we have $\chi(\mathbb{U} \otimes_u \mathbb{W}) = 0$. Then by (3), $\chi(\mathbb{U} \oplus_u \mathbb{W}) \notin span\{\chi(\mathbb{U}), \mathbb{U} \in \mathcal{H}\}$. Moreover, by Claim 4.1, we have $p(\mathbb{U} \oplus_u \mathbb{W}) > 0$. Thus Claim 6.3 is applicable, and it contradicts the choice of $\mathbb{W}$ as an eligible set-pair that crosses the minimum number of set-pairs in $\mathcal{H}$. An analogous argument shows $p(\mathbb{U} \oplus_u \mathbb{W}) > 0$.

Since $\chi(\mathbb{W}) \notin span\{\chi(\mathbb{U}), \mathbb{U} \in \mathcal{H}\}$, (3) implies that either $\chi(\mathbb{U} \otimes_u \mathbb{W})$ or $\chi(\mathbb{U} \oplus_u \mathbb{W})$ is also not contained in this set. Again we can use Claim 6.3 to derive a contradiction. This completes the proof of Lemma 6.1. □